\documentstyle[preprint,tighten,aps]{revtex}
\begin{document}
\draft
\preprint{DAMTP/R/96/31,QMW-96-12,hep-th/9606193}
\date{June 1996}
\title{Black Holes and U-Duality}
\author{Mirjam Cveti\v c$^1$
\thanks{On sabbatic leave from the University of Pennsylvania.
}
and Christopher M. Hull$^2$
}
\address{$^1$ School of Natural Science, Institute for Advanced
Study,\\ Olden Lane, Princeton, NJ 08540, U.S.A.,\\ $\ $ \\
$^1$ Department of Applied Mathematics and Theoretical Physics,\\University
of Cambridge, Cambridge CB3 9EW, U.K. \\ and\\
$^2$  Physics Department, Queen Mary and Westfield College, \\
 Mile End Road, London E1 4NS, U.K.}
\maketitle
\begin{abstract}
{We find  the general charged rotating black hole
solutions of the maximal supergravities in  dimensions $4\le D\le 9$
arising from toroidally compactified   Type II string or M-theories.
In each dimension,
these are obtained by acting on a generating solution with classical duality
symmetries. In $D=4$, $D=5$ and $6\le D \le 9$ the generating solution
is specified by the ADM mass,  $[{{D-1}\over 2}]$-angular momentum
components
and  five, three  and two charges, respectively.
We discuss the  BPS-saturated  (static) black holes
and derive the $U$-duality invariant form of the area
of the horizon. We also comment on the  $U$-duality
invariant form of the
BPS     mass formulae.
}
\end{abstract}

\section{Introduction}

Black holes play an important role in string theory and
recent developments  (for a   review, see
 \cite{Horowitz}) have shown that string theory makes it possible  to address
 their microscopic  properties,
in particular the statistical origin of the  black hole
 entropy and possibly  issues of information loss.
The starting point in such investigations is the classical black hole solution
and the aim of this paper is to find such solutions in toroidally compactified
string theories  in dimensions $4\le D \le 9$.

The general solutions are found from a particular black  hole  ``generating
solution'', which is specified by a  canonical choice of the
asymptotic values of the scalar fields, the ADM  mass,   $[{{D-1}\over
2}]$-components of angular
momentum
and  a  (minimal) number of charge parameters.
The most general black hole,
 compatible with the ``no-hair theorem'',
 is   then  obtained by acting on   the generating solution  with classical
duality  transformations.
These are symmetries of the supergravity equations of motion, and so generate
new solutions from old.
They  do
not change
the $D$-dimensional Einstein-frame metric but do change the charges and scalar
fields.
We first  consider transformations,  belonging to the maximal compact subgroup
of duality transformations,  which  preserve the canonical
asymptotic values of the scalar fields and show that all charges are generated
in this way.
Another  duality transformation can be used to change the
asymptotic values of the scalar fields.

For  the  toroidally compactified heterotic string such a program is now close
to completion.
Particular examples of solutions had been obtained in a number of papers  (for
a recent review and references, see
 \cite{Horowitz}).
In dimensions $D=4$, $D=5$ and $6\le
D\le 9$ the generating solution   has
five,  three and two charge parameters, respectively.
The charge parameters of the generating solution are   associated with
the $U(1)$  gauge fields arising from  Kaluza-Klein (momentum modes)    and
two-form (winding modes)  sectors
 for at most two toroidally
compactified directions. The general black hole solution is then
obtained by applying to the generating solution a
subset of transformations, belonging to  the
maximal compact subgroup  of the  $T$- and $S$-duality transformations
\cite{SEN}.
The explicit  expression for the  generating
solution  has  been obtained in
 $D=5$ \cite{CY5r}
and $D\ge 6$\cite{CYNear,Llatas}, however,
 in $D=4$ only  the five charge  static  generating solution
\cite{CY4s} (see also \cite{JMP})
 and the four
charge rotating solutions \cite{CY4r} were obtained.

The BPS-saturated
solutions of  the toroidally compactified  heterotic string
 have non-singular horizons only  for   $D=4$ static
black holes  \cite{CY,CTII}  and   $D=5$ black holes with one non-zero
angular momentum component \cite{TMpl}.  In $6\le D\le 9$
 the BPS-saturated black
holes have  singular horizons  with  zero area.
The explicit $T$- and
$S$-duality invariant  formulae for
 the area of the horizon and the
ADM mass for the general  BPS-saturated black holes were   given for
$D=4$ in \cite{CY,CTII} and for  $D=5$ in \cite{TMpl}. In
particular,
 the area of the horizon of the  BPS-saturated
black holes  {\it does not} depend  on the asymptotic values of the scalar
fields \cite{LWI,CTII,S,FK}, suggesting   a  microscopic interpretation.

The purpose of this paper is to   study properties of the
classical black hole solutions  of {\it toroidally compactified
 Type II string  theory or ${\rm M}$-theory},
 in dimensions $4\le D\le 9$, thus completing the program for the
toroidally compactified superstring vacua.
  We  identify the
minimum number of  charge parameters for the generating
solutions,  which fully specifies the  space-time metric of  the {\it general}
black hole in $D$-dimensions.  The
``toroidally''  compactified  sector of the  heterotic string and the
Neveu-Schwarz-Neveu-Schwarz   (NS-NS) sector of  the toroidally
compactified Type  II  string have the {\it same} effective action and
so  have the same     classical solutions.  In this paper we shall show  that
the generating
solutions  for black
holes  in the toroidally compactified Type II string
 theory (or M-theory) are  the {\it same} as the ones  of
toroidally compactified heterotic string. Note that it could have been the case
that a more general generating solution with one or more RR charges was needed.
Applying $U$-duality  transformations  to the generating solution generates
all black  holes of
toroidally  compactified Type II string theory (or M-theory).

 We further address
 the   BPS-saturated solutions,
identify  the  $U$-duality invariant
 expression for the area of the horizon, i.e. Bekenstein-Hawking (BH) entropy
for the general  BPS-saturated black holes
 and outline a
procedure to obtain the  manifestly $U$-duality invariant
mass formulae.

The paper is organized as follows.
In Section II we  summarize the  symmetries of the
effective action of toroidally compactified
 Type II string
and  obtain  the general solution  by
applying a   compact subgroup
of  duality transformations to the generating solution.
In Section III  we  concentrate on  general   static BPS-saturated
black holes in $4\le D\le 9$  and  derive the $U$-duality invariant
  expression for the area of the
horizon.
In Appendix A the effective action of
 the NS-NS sector
of toroidally compactified Type II string is given.
The explicit form of some of the generating solutions is given in
in Appendix B.

\section{ Toroidally Compactified Type II
String Theory }

\subsection{Symmetries}

The low-energy effective action for
the Type II string  or M-theory, toroidally compactified
 to $D$-dimensions,  is
the maximal supergravity theory, which has a continuous duality symmetry $U$ of
its equations of motion \cite{CJ}  (see Table I, first column). This  has a
maximal compact subgroup $C_U$  (second  column in  Table I).  In the  quantum
theory
the continuous classical symmetry $U$
is broken to  a discrete  subgroup ${ Q}_{ U}$  \cite{HT} (third  column in
Table I) which is the  $U$-duality symmetry of the string theory. However, we
will sometimes refer to the group $U$ as the classical  $U$-duality.

\subsection{Solution Generating Technique}

The general black hole solution is obtained by acting on  generating
solutions   with   $U$-duality transformations.

The scalar fields  take values in the coset $U/C_U$ and
can be parameterised
by a $U$-valued matrix
${\cal V}(x)$ which transforms under rigid $U$-transformations from the right
and local $C_U$
transformations from the left \cite{CJ}.  The  Kaluza-Klein and
 and antisymmetric
tensor  $U(1)$ gauge fields
also transform under $U$. It is convenient to define ${\cal M}={\cal V}^t{\cal
V}$ which is inert
under $C_U$ and transforms under $U$ as
${\cal M}{} \to { \Omega} {\cal M}{}{
\Omega}^T $ (${\Omega} \in { U}$).

The asymptotic value ${\cal
M}_{\infty}  $ of
${\cal M}$ can be brought to the canonical value ${\cal M}_{0\, \infty}= {\bf
1}$ by a suitable $U$-duality transformation $\Omega_0$. The canonical value
${\cal M}_{0\, \infty}$ is
preserved by
 $C_U$ and the most general solution with the asymptotic behaviour ${\cal
M}_{\infty} ={\cal
M}_{0\, \infty}$ is obtained by acting on the generating solution with a subset
of $C_U$ transformations, i.e. the $C_U$ orbits which are of
the form $C_U/C_0$ where $C_0$ is the subgroup preserving the generating
solution.  In particular, with this procedure
the complete set of charges is obtained.  Indeed,    the generating solution is
labelled by $n_0$ charges
($n_0=5,3,2$ for $D=4,5,\ge 6$,  respectively) and   if  the dimension of the
$C_U$
orbits  is
$n_1$, then
$n_0+n_1$ is the correct dimension of the vector space of charges for the
general solution, as we shall
check in the following Section. Black holes with
arbitrary asymptotic values of scalar fields  ${\cal M}_{\infty}$ can then be
obtained
from these by acting
with $\Omega_0$.

We shall seek the general charged rotating black hole
solutions.
In $D=4$,    such  solutions are specified by electric {\it and}  magnetic
charges, while in $D>4$ they carry electric charges only (once all $(D-3)$-form
gauge fields  have been dualised to vector fields).

\section{Black Holes in Various Dimensions}

We will first propose generating solutions for Type II string (or M-theory)
black holes in
dimensions $4\le D\le
9$ and then go on to show that the action of duality transformations generates
all solutions.
Remarkably, the generating solutions are the same as those used for the
heterotic string. We will
discuss only the charge assignments here, and give the explicit solutions in
Appendix B.

\subsection{Charge Assignments for the Generating Solution}\label{cgs}
\subsubsection{D=4}

The generating solution is specified in terms of {\it five}
charge parameters.
It is convenient to choose these to arise in
the NS-NS sector of the
compactified Type II  string  as follows.
We choose two  of the toroidal dimensions labelled by $i=1,2$ and let $A_{\mu\,
i}^{(1)}$ be the
two graviphotons (corresponding to $G_{\mu i}$) and $A_{\mu\, i}^{(2)}$ the two
$U(1)$ gauge fields coming
from the antisymmetric tensor (corresponding to $B_{\mu i}$) (see Appendix A).
Corresponding to
these four  $U(1)$ gauge
fields there are four electric charges $Q^{(1),(2)}_i$ and four magnetic ones
$P^{(1),(2)}_i$. The
generating solution, however  carries the following five charges:
$ Q_1\equiv Q_1^{(1)},\ Q_2\equiv Q_1^{(2)}, \ P_1\equiv P_2^{(1)},
P_2\equiv P_2^{(2)}$  and
$q\equiv Q_2^{(1)}=-Q_2^{(2)}$.   It will be useful to define the
left-moving and right-moving charges
 $Q_{i\, L,R}\equiv Q_i^{(1)}\mp Q_i^{(2)}$ and
$P_{i\, L,R}\equiv P_i^{(1)}\mp P_i^{(2)}$ ($i=1,2$).
The generating
 solution   then carries five
charges   associated with the first two compactified toroidal
 directions of the NS-NS sector, where  the
dyonic charges are subject to the constraint ${\vec P}_R{\vec Q}_R=0$.
 We choose the
convention that all the five charge parameters are positive.

\subsubsection{D=5}
In $D=5$ the generating solution is  parameterised  by three
(electric)  charge
parameters:
$Q_1\equiv Q_1^{(1)}, \ Q_2\equiv Q_1^{(2)},$ and  $ {\tilde Q}$. Here
the electric charges $Q_i^{(1),(2)}$ arise
respectively from  the graviphoton $A_{\mu\, i}^{(1)}$
 and antisymmetric tensor
$A_{\mu\, i}^{(2)}$  $U(1)$ gauge
 fields of the $i$-$th$ toroidally compactified direction of the
NS-NS sector,
 and $\tilde Q$ is the  electric charge  of the gauge field, whose field
strength is related to the field strength of the two-form field $B_{\mu\nu}$
by
duality transformation (see Appendix A). Again we choose the
convention that all three charges are positive.

\subsubsection{$6\le D\le 9$}
In $6\le D\le 9$ the generating solution is  parameterised  by two
electric  charges:
$Q_1\equiv Q_1^{(1)}, \ Q_2\equiv Q_1^{(2)}$. Again,
the electric charges $Q_i^{(1),(2)}$ arise
respectively from  the graviphoton $A_{\mu\, i}^{(1)}$
 and antisymmetric  tensor
$A_{\mu\, i}^{(2)}$ $U(1)$ gauge
 fields of the $i$-$th$  toroidally compactified direction and we use the
convention that both charges are positive.

Note that the explicit form of the generating solutions
 with the  above charge assignments is the {\it same} as the one  of the
toroidally compactified heterotic string, since the corresponding
NS-NS sector of the toroidally compactified string and the
``toroidal'' sector of the heterotic string are the same.

\subsection{Action of Duality Transformations   on  Generating Solution}
\subsubsection{D=4}
The $N=8$ supergravity has 28 abelian gauge fields and so the  general black
hole solution carries
56 charges (28 electric and 28 magnetic). The  ${ U}$-duality group is
$E_{7(7)}$, the maximal compact subgroup ${ C_U}$ is $SU(8)$
and the $T$-duality subgroup is $SO(6,6)$ . We use the formulation with rigid
$E_7$ symmetry and
local $SU(8)$ symmetry \cite{CJ}. The 56 charges fit into a vector ${\cal Z}$
transforming as a {\bf
56} of
$E_7$. In the quantum theory, ${\cal Z}$ is constrained to lie in a lattice
by charge
quantisation
\cite{HT}. This ``bare'' charge vector can be ``dressed'' with the asymptotic
value ${\cal V}_\infty$ of
the  scalar field  matrix ${\cal V}$ to form
 \begin{equation}
\bar {\cal Z}={\cal V}_\infty{\cal Z}=
\left ( \matrix{
q^{ab}\cr
p_{ab}\cr}
\right ) \ ,
\label{zv}
\end{equation}
which is invariant under $E_7$ but transforms under local $SU(8)$.
The   28 electric and  28 magnetic dressed charges are
 $q_{ab}$ and $p_{ab}$  ($a,b=1, \cdots , 8$ and
$q_{ab}=-q_{ba}$,$p_{ab}=-p_{ba}$).
 They
can be combined to form the
$Z_{4\,AB}$ matrix ($A,B=1, \cdots , 8$ are $SU(8)$ indices)
 transforming as  the complex antisymmetric
 representation of $SU(8)$, by defining $
Z_{4\,AB}= (q^{ab}+ip_{ab}) (\Gamma^{ab})^B{}_A$ where $(\Gamma^{ab})^B{}_A$
are
the generators of $SO(8)$ in the spinor representation  \cite{CJ}. The matrix
$Z_{4\,AB}$
appears on the right
hand side of the
anticommutator of   chiral two-component supercharges
  \begin{equation}
[Q _{A\alpha}, Q_{B\beta}]= C_{\alpha\beta} Z_{4\,AB} \ ,
\end{equation}
and thus corresponds to the  matrix of 28 complex central charges.
 An $SU(8)$ transformation
$ Z_4\rightarrow  Z^0_4=({\cal U} Z_4{\cal  U}^T)$
  brings this charge matrix   to the skew-diagonal form:
  \begin{equation}
Z^0_4= \left ( \matrix{
0& \lambda_1&0&0&0&0&0&0 \cr
-\lambda_1&0&0&0&0&0&0&0 \cr
0&0&0& \lambda_2&0&0&0&0 \cr
0&0&-\lambda_2&0&0&0&0&0\cr
0&0&0&0&0& \lambda_3&0&0\cr
0&0&0&0&-\lambda_3&0&0&0\cr
0&0&0&0&0&0&0& \lambda_4\cr
0&0&0&0&0&0&-\lambda_4&0\cr}
\right )\ ,
\label{diag4}
\end{equation}
where the complex $\lambda_{i}$ ($i=1,2,3,4$) are the skew eigenvalues.

For the generating solution with
the five  charge parameters  $Q_{1,2}$, $P_{1,2}$
and $q$
(see Subsection \ref{cgs}) the eigenvalues are \footnote{For
the four charges $Q_{1,2}$ and $P_{1,2}$  the eigenvalues were
given in \cite{KK}.}:
\begin{eqnarray}
\lambda_1&=&Q_{1R} + P_{2R},\cr
 \lambda_2&=&Q_{1R} -P_{2R},\cr
 \lambda_3&=&Q_{1L} + P_{2L}+2iq,\cr
 \lambda_4&=&Q_{1L} - P_{2L}-2iq
 \label{fivepar}
\end{eqnarray}
(recall $Q_{1\, L,R}\equiv Q_1\mp Q_2$ and
$P_{2\, L,R}\equiv P_1\mp P_2$).

We now consider the action of duality transformations  on the generating
solution
and show that all $D=4$ black hole solutions  are indeed generated.
The $U$-duality group $E_7$ has a maximal  subgroup $SO(6,6)\times SL(2,{\bf
R})$
where  $SO(6,6)$ is the $T$-duality group and $ SL(2,{\bf R})$ is the
$S$-duality
group.
(Strictly speaking, the duality groups are discrete subgroups of these.)
The {\bf 56} representation of $E_7$ decomposes as
  \begin{equation}
{\bf 56} \to (12,2) \oplus (32,1)
  \end{equation}
under $SO(6,6)\times SL(2,{\bf R})$ and  thus the 56 charges ${\cal Z}$
decompose
into $12$ electric and
$12$ magnetic charges in the NS-NS sector, and 32 charges in the  Ramond-Ramond
(RR) sector.
We choose for now the  asymptotic value of the scalars to be the canonical one,
i.e.
${\cal V}_\infty={\cal V}_{0\infty}\equiv {\bf
1}$.

The maximal compact symmetry of the $T$-duality group is
$SO(6)_L\times SO(6)_R\sim SU(4)_L\times SU(4)_R$, and under $ SU(4)_L\times
SU(4)_R\subset SU(8)$
the complex representation ${\bf 28}$ decomposes into  complex representations
$(12,1)+(1,12)+(4,4)$. This decomposition corresponds to
splitting the
 $8\times 8$ matrix of charges
$Z_4$ into $4\times 4$ blocks. The
two  $4\times 4$
diagonal blocks $Z_R$ and $Z_L$,   transform  respectively as the
antisymmetric
complex representations of $SU(4)_{R,L}\sim SO(6)_{R,L}$ and represent the
$12+12$ charges of the
NS-NS sector. The off-diagonal blocks correspond to the 16 complex RR charges.

The maximal compact subgroup of $SO(6,6)\times SL(2,{\bf R})$ is $SO(6)_L\times
SO(6)_R\times
SO(2)$ and it preserves ${\cal V}_{0\infty}$.
The subgroup that preserves the charges of the generating solution
is $SO(4)_L\times SO(4)_R$. Thus   acting on the generating solution with
$SO(6)_L\times SO(6)_R\times
SO(2)$, gives orbits corresponding to  the 19-dimensional space
\begin{equation}
{SO(6)_L\times SO(6)_R \over SO(4)_L\times SO(4)_R} \times SO(2)\ .
\label{osp}
\end{equation}
As the generating solution has five charges, acting on the generating solution
with $SO(6)_L\times
 SO(6)_R\times
SO(2)$
 gives the
required
$5+19=24$  NS-NS charges, i.e.  the 24  NS-NS charges are parameterised in
terms of the five
charges of the
generating solution and the 19 coordinates of the orbit space (\ref{osp}).

The above procedure is closely related to that for  $D=4$  toroidally
compactified heterotic string vacua \cite{CY,CTII},
where the
general black hole  with the  $5+51=56$
charges is obtained from the same five-parameter generating solution,
and the 51
coordinates of
 the orbit
\begin{equation}
{SO(22)_L\times SO(6)_R\over SO(20)_L\times SO(4)_R} \times SO(2) \ .
\end{equation}

We  can now   generalise this procedure  to include the RR charges.
The group $C_U=SU(8)$ preserves the canonical asymptotic values of the scalar
fields
and only the  subgroup   $SO(4)_{L}\times SO(4)_{R}$
leaves the generating solution invariant.
Then acting with $SU(8)$ gives orbits
\begin{equation}
SU(8)/[SO(4)_L\times SO(4)_R]
\label{4du}
\end{equation}
of dimension  $63-6-6=51$. The $SU(8)$ action then  induces  $51$  new charge
parameters,
 which along with the original
five parameters  provide charge parameters for the general
 solution with 56 charges.
Finally, the general black hole with arbitrary asymptotic values of the scalars
is obtained from
these 56-parameter solutions by acting with a $E_7$ transformation. This
transformation leaves
the central charge
matrix $Z_4$  and its eigenvalues $\lambda_i$ invariant, but changes the
asymptotic values of
scalars and ``dresses'' the physical charges.  The orbits under
$E_7$ are the 70-dimensional coset $E_7/SU(8)$, as  expected.

The fact that the same five-parameter generating solution that was used for the
$D=4$ toroidally compactified heterotic string
should be sufficient to generate all black holes  with NS-NS charges of  $D=4$
toroidally compactified  Type II  is unsurprising, given the equivalence
between the  ``toroidal'' sector  of the heterotic
string and the NS-NS sector of the Type II string. However, it is interesting
that the procedure outlined above  is also
sufficient to generate {\it all} RR charges of the general black hole solution,
as it could have been the case that a
more general generating
solution carrying one or more RR charges was needed.

\subsubsection{D=5}

The  ${ U}$-duality group is  $E_{6(6)}$, the maximal compact subgroup ${ C_U}$
is $USp(8)$ and
the $T$-duality group is $SO(5,5)$ with its maximal compact subgroup
$SO(5)_L\times SO(5)_R$.  In this case there are 27 abelian gauge
fields and the 27 electric
charges (dressed with  asymptotic values of the scalar fields) transform as a
{\bf 27} of USp(8) and can be
assembled into an
$8\times 8$ matrix
 $Z_{5\,AB}$  ($A,B=1,\dots, 8$) with the properties
\cite{Cremmer}:
\begin{equation}
Z_5^{AB\, *}=\Omega^{AC}\Omega^{BD} Z_{5\,CD}, \ \  \Omega^{AB}Z_{5\,AB}=0  ,
\label{Z5}
\end{equation}
where $\Omega$ is the  $USp(8)$ symplectic invariant, which we take to be
\begin{eqnarray}
\Omega= \left ( \matrix{
0&1&0&0&0&0&0&0 \cr
-1&0&0&0&0&0&0&0 \cr
0&0&0&1&0&0&0&0 \cr
0&0&-1&0&0&0&0&0\cr
0&0&0&0&0& 1&0&0\cr
0&0&0&0&-1&0&0&0\cr
0&0&0&0&0&0&0& 1\cr
0&0&0&0&0&0&-1&0\cr}
\right ) .
\label{explo}\end{eqnarray}
With $\Omega$ given by (\ref{explo}), the $Z_5$  charge
matrix can be written in  the  following form:
\begin{eqnarray}
Z_5= \left ( \matrix{
0& z_{12}& z_{13}&z_{14}& \cdots \cr
-z_{12}&0&-z_{14}^*&z_{13}^*&\cdots \cr
-z_{13}^*&z_{14}&0&z_{34}&\cdots\cr
-z_{14}^*&-z_{13}&-z_{34}&0&\cdots\cr
\cdots&\cdots&\cdots&\cdots&\cdots}
\right ).
\label{explZ0}\end{eqnarray}
 Here $z_{12},\ z_{34}, \ z_{56}$ are real and satisfy $ z_{12}+ z_{34}+
z_{56}=0$.

The matrix $Z_5$ occurs in the superalgebra and represents the 27 (real)
central
charges.
It can be brought into a skew-diagonal form  of the
type (\ref{diag4}) by
an  $USp(8)$ transformation
$ Z_5\rightarrow  Z^0_5=({\cal U} Z_5 {\cal U}^T)$. The  four real eigenvalues
$\lambda_i$
are subject to the  constraint: $\sum_{i=1}^4\lambda_i=0$.

The generating solution is  parameterised  by three charges
$Q_1\equiv Q_1^{(1)},\ Q_2\equiv Q_1^{(2)}$ and  $\tilde Q$ (see Subsection
\ref{cgs}).
The four (constrained, real) eigenvalues $\lambda_i$ are then
\begin{eqnarray}
\lambda_1&=& \tilde Q+Q_{1R},\cr
 \lambda_2&=&\tilde Q-Q_{1R},\cr
 \lambda_3&=&-\tilde Q+Q_{1L},\cr
 \lambda_4&=&-\tilde Q-Q_{1L},
 \label{threepar}
\end{eqnarray}
where
 $Q_{1\, L,R}\equiv Q_1\mp Q_2$. These indeed satisfy the   constraint
$\sum_{i=1}^4\lambda_i=0$.

The  three  parameter solution  is indeed the  generating
solution for black holes in $D=5$. The $USp(8)$ duality transformations
preserve the canonical
asymptotic values of the scalar fields and the subgroup    $SO(4)_L\times
SO(4)_R[\subset
SO(5)_L\times SO(5)_R]
\subset USp(8)$
preserves the generating solution.
Acting with $USp(8)$ on the generating solution gives orbits
\begin{equation}
USp(8)/[SO(4)_L\times SO(4)_R]\ ,
\label{5du}
\end{equation}
and thus  induces  $36 -4\times 3=24$
  new charge parameters,
 which along with the original
three charge parameters  provide  27 electric charges for
  the general   solution  in $D=5$.

\subsubsection{D=6}

The  ${ U}$-duality group is  $SO(5,5)$, the maximal compact subgroup ${ C_U}$
is $SO(5)\times
SO(5)$ and the $T$-duality group is $SO(4,4)$ which has  maximal compact
subgroup $SO(4)_L\times SO(4)_R\sim [SU(2)\times SU(2)]_L\times [SU(2)\times
SU(2)]_R
$.  There are 16 abelian vector
fields and the
bare charges ${\cal Z}$ transform as a {\bf 16} (spinor)   of $SO(5,5)$.
The dressed charges transform as  the $(4,4)$ representation of $SO(5)\times
SO(5)$ and can be
arranged into a
$4 \times 4 $ charge matrix $Z_6$.

Under $ [SU(2)\times SU(2)]_L\times [SU(2)\times SU(2)]_R\subset SO(5)\times
SO(5)$
the $(4,4)$ decomposes into $(2,2,1,1) +(1,1,2,2)+(1,2,2,1)+(2,1,1,2)$. This
decomposition corresponds to
splitting the
 $4\times 4$ matrix of charges
$Z_6$ into $2\times 2$ blocks. The
two  $2\times 2$
diagonal blocks $Z_R$ and $Z_L$,   transform  respectively as  $(2,2,1,1)$ and
$(1,1,2,2)$ representations  of $ [SU(2)\times SU(2)]_L\times [SU(2)\times
SU(2)]_R$  representing  the
$4+4$ charges of the
NS-NS sector. The off-diagonal blocks correspond  to  $(1,2,2,1)+(2,1,1,2)$
representations of $ [SU(2)\times SU(2)]_L\times [SU(2)\times SU(2)]_R $ and
represent
8  RR charges.

The matrix $Z_6$ occurs in the superalgebra and represents the 16 (real)
central
charges.
It can be brought into a skew-diagonal form  of the
type (\ref{diag4}) by
an  $SO(5)\times SO(5)$ transformation
$ Z_6\rightarrow  Z^0_6=({\cal U} Z_6 {\cal U}^T)$ with the  two eigenvalues
$\lambda_i$.
The generating solution is  parameterised  by two charges
$Q_1\equiv Q_1^{(1)}, \  Q_2=Q_1^{(2)}$ (see Subsection \ref{cgs}).
The two  eigenvalues $\lambda_i$ are then
\begin{eqnarray}
\lambda_1&=& Q_{1R},\cr
 \lambda_2&=&Q_{1L},
 \label{twopar}
\end{eqnarray}
where again
 $Q_{1\, L,R}\equiv Q_1\mp Q_2$.

The generating solution is preserved by   $SO(3)_L\times SO(3)_R [\subset
 SO(4)_L\times SO(4)_R]\subset SO(5)\times SO(5)$
so acting with
  $SO(5)\times SO(5)$  gives
\begin{equation}
[SO(5)\times
SO(5)]/[SO(3)_L\times SO(3)_R]
\label{6dtr}
\end{equation}
 orbits, and thus introduces  $2 (10-3)=14$ charge parameters,
which along  with the  two charges ($Q_{1,2}$)
of the generating solution  provide the
16 charge parameters of the general solution in $D=6$.

\subsubsection{D=7}

The  ${ U}$-duality group is  $SL(5,{\bf R})$,
the maximal compact subgroup ${ C_U}$ is $SO(5)$ and  the $T$-duality group is
$SO(3,3)$ with its maximal compact subgroup $SO(3)_L\times SO(3)_R$.
There are ten abelian vector fields and the ten  bare electric charges
transform as the $\bf 10$ representation of $SL(5, {\bf  R})$. Dressing of
these
with  asymptotic values of scalars gives the
ten central charges which are inert under $SL(5, {\bf  R})$ but transform as a
$\bf 10$ under
$SO(5)$.  The dressed charges can be assembled into a real antisymmetric
$5\times 5$ charge matrix  $Z_{7ij}$, which appears in the superalgebra as the
$4 \times 4$ central charge matrix
$Z_{7AB}={1 \over 2}Z_{7ij}{\gamma ^{ij}}_{AB}$ where $ {\gamma ^{ij}}_{AB}$
are the
generators of $SO(5)$ in the spinor ({\bf 4}) representation. The matrix
$Z_{7ij}$ has two real skew eigenvalues, $\lambda_1, \lambda_2$, which  for the
generating solution correspond to
the two charges $Q_{1\, L,R}$.

The  subgroup $SO(2)_L\times SO(2)_R\subset SO(3)_L\times SO(3)_R\subset
O(5)$
preserves the generating solution, so that the action of $SO(5)$ gives orbits
\begin{equation}
SO(5)/[SO(2)_L\times SO(2)_R]\ ,
\label{6du}
\end{equation}
thus introducing
 $10-2=8$ charge parameters,
which together with the  two charges ($Q_{1,2}$) of the generating
solution (see Subsection \ref{cgs})
provide the
ten   charges of the general solution in $D=7$.

\subsubsection{D=8}

The  ${ U}$-duality group is  $SL(3,{\bf R})\times SL(2, {\bf R})$,
the maximal compact subgroup ${C_U}$ is $SO(3)\times U(1)$ and  the $T$-duality
group is $SO(2,2)$ with   maximal compact subgroup $SO(2)_L\times SO(2)_R$.
There are  six abelian gauge fields and the six bare electric charges transform
as
$({\bf 3},{\bf 2})$
under $SL(3, {\bf  R})\times SL(2,{\bf R})$.

No $C_U$ transformations preserve the generating solution, so that the orbits
are
 \begin{equation}
C_U=SO(3)\times U(1)\ ,
\label{7du}
\end{equation}
and
   the $C_U$ transformations  introduce  $(3+1)$-charge
parameters, which along with the  two charges  of the generating solution
provide the
six  charges of the general solution in $D=8$.

\subsubsection{D=9}

The  ${U}$-duality group is  $SL(2,{\bf R})\times {\bf R}^+$ and
the maximal compact subgroup ${ C_U}$ is $U(1)$.
There are three abelian gauge fields and the  three bare  electric charges
transform
as
$({\bf 3}, {\bf 1})$ under $SL(2,{\bf R})\times {\bf R}^+$.
 The action of
\begin{equation}
C_U=U(1),
\label{9du}
\end{equation}
   introduces  one-charge
parameter, which along with the  two charges  of the generating
solution  provides  the
three charges of the general solution in $D=9$.

 \section{Entropy and Mass of BPS-Saturated Static Black Holes}

We now  study  the properties of
static BPS-saturated solutions. In the preceding Section
 we   identified the  charge assignments for the
 generating solutions, which fully specify the space-time of the  general
black hole solution  in $D$-dimensions for toroidally compactified
Type II  string  (or M-theory) vacuum.
The explicit form for  these solutions
 has been given in the literature (with the exception of the rotating
five-charge solution in $D=4$).
The static generating solutions are given in Appendix B.
Also, the area of the horizon for the BPS-saturated (as well as for
non-extreme
solutions) was calculated explicitly.
In addition to static BPS-saturated solutions, we shall also consider
near-BPS-saturated solutions
in
$4\le D\le 9$.

\subsection{The Bogomol'nyi Bound}

Consider first the $D=4$ case.
Standard arguments \cite{GH} imply that the ADM mass $M$ is bounded below by
the
moduli of the
eigenvalues $\lambda _i$  (\ref{diag4}) of the  central charge matrix  $Z_4$,
i.e.
$M \ge |\lambda _i|$, $i=1,...,4$. Without loss of generality the eigenvalues
can be ordered in  such a way that  $|\lambda_i|\ge |\lambda_j|$ for $j\ge i$ .

If $M$ is equal to $|\lambda _1|=\cdots =|\lambda_p|$,
 the solution preserves
${p\over 8}$ of $N=8$ supersymmetry\footnote{For  a related discussion  of the
number of preserved supersymmetries see  \cite{Kal,KK}.}. For example, if
$M=|\lambda
_1|>|\lambda
_{2,3,4}|$ then ${\textstyle{1\over 8}}$ of the supersymmetry is preserved,
while for
$M=|\lambda _1|=|\lambda _2|=|\lambda _3|=|\lambda _4|$,  ${\textstyle{1\over
2}}$ of the
supersymmetry is
preserved.

The eigenvalues $\lambda _i$ are each invariant under $E_7$ and $SU(8)$.
The physical quantities such as  the Bekenstein-Hawking (BH) entropy and the
ADM mass of  BPS-saturated black holes, can then be
written in terms of these
quantities, which depend on both the bare charges $\cal Z$ and the  asymptotic
values of the scalar fields parameterised by ${\cal V}_{\infty}$. However,
there are special combinations of these invariants for which the dependence on
the  asymptotic values of scalar fields drops out. In particular, such
combinations   play a special role in the  BH entropy for the BPS-saturated
black hole solutions as discussed in the following Subsection.
Similar comments apply in $D>4$.

\subsection{Bekenstein-Hawking Entropy}
 The BH entropy is defined  as
 $S_{BH}= {\textstyle{1\over {4G_D}}} A_{h}$ where  $G_D$ is the
$D$-dimensional Newton's constant
and $A_{h}$  is the area of the horizon.  Since the Einstein metric is duality
invariant,  geometrical
quantities  such as $A_{h}$ should be too.
Thus it should be possible to write  $A_{h}$ in terms of duality invariant
quantities such as
the eigenvalues $\lambda _i$, the ADM mass (or the ``non-extremality''
parameter $m$ defined in
Appendix B) and the angular momentum components.  However, in the case of
BPS-saturated black holes  the BH entropy is, in addition, {\it independent }
of the  asymptotic values of scalar fields. This  property
was pointed  out in \cite{LWI},  and further exhibited
for the general BPS-saturated solutions  of toroidally compactified
heterotic string vacua in  $D=4$ \cite{CTII}  and $D=5$ \cite{CY5r} as well as
for  certain
BPS-saturated black holes of  $D=4$ $N=2$ superstring vacua
in \cite{FKS,S}. This property was explained in terms of ``supersymmetric
attractors''
in \cite{FK}.

When applying   the above arguments to  BPS-saturated  black holes of
toroidally compactified Type II string (or M-theory)  their  BH entropy  can
be written in terms of a $U$-duality invariant combination of {\it bare}
charges
alone,   thus implying  that  only  a very special combination of bare charges
can appear in the BH entropy formula.
One is then led to the remarkable result that the entropy must be
given in terms of the
quartic invariant of $E_7$ in $D=4$ and the cubic invariant of $E_6$ in $D=5$,
as these are the
only possible $U$-invariants of bare charges.\footnote{In $D=3$ there is a
unique quintic  $E_{8(8)}$-invariant which should play a similar role
for $D=3$ black hole solutions.}
This fact was first pointed out in \cite{KK} and \cite{FK}, respectively (and
independently
in \cite{CHU}). It  has been checked explicitly for  certain classes of $D=4$
BPS-saturated black
holes
\cite{KK}. Below we will extend the analysis to general BPS-saturated black
holes in
$D=4,5$.

 For $6\le D\le 9$
 there are {\it no} non-trivial
$U$-invariant quantities that can be constructed from the bare charges alone,
in agreement with the result
that there are  no
BPS-saturated black holes with non-singular horizons and   finite BH entropy in
$D\ge 6$,   as has  been
shown explicitly in
\cite{KT,CYNear}.

\subsubsection{D=4}
The five-parameter static generating solution has the following BH entropy
\cite{CTII}:
\begin{equation}
S_{BH}= 2\pi\sqrt{Q_1Q_2P_1P_2-{\textstyle{1\over 4}}q^2(P_1+P_2)^2}.
\label{4ent}
\end{equation}
We shall now show that (\ref{4ent})  can be rewritten in terms of the $E_7$
quartic invariant (of bare charges).

The  quartic  $E_{7(7)}$  invariant
    $J_4$, constructed from the charge matrix $Z_{4\,AB}$, is
\cite{CJ}:
\begin{eqnarray}
{J_4}&= &{\rm Tr}({ Z_4}^\dagger{ Z_4})^2-{\textstyle{1\over 4}}
({\rm Tr}Z_4^\dagger{
Z_4})^2+\cr
& &{\textstyle {1\over {96}}}(\epsilon_{ABCDEFGH}{ Z_4}^{AB\, *}{
Z_4}^{CD\, *}{
Z_4}^{EF\, *}
{ Z_4}^{GH\, *}+\epsilon^{ABCDEFGH}Z_{4\, AB}Z_{4\, CD} Z_{4\, EF} Z_{4\, GH})\
,
\label{quartic}
\end{eqnarray}
which  can be written in terms of the skew-eigenvalues $\lambda _i$ by
substituting the
skew-diagonalised   matrix $Z^0_{4}$ (\ref{diag4}) in (\ref{quartic}) to give
(as in  \cite{KK}):
\begin{eqnarray}
{J_4}&=&\sum_{i=1}^4 |\lambda_i|^4 -2\sum_{j>i=1}^4
|\lambda_i|^2|\lambda_j|^2\cr
&+&4({ \lambda}_1^*{\lambda}_2^*{\lambda}_3^*{ \lambda}_4^*+
\lambda_1\lambda_2\lambda_3\lambda_4)\ .
\label{diagfour}
\end{eqnarray}
For the five parameter generating  solution, the $\lambda _i$ are given by
(\ref{fivepar}), so that
  (\ref{diagfour}) becomes:
\begin{eqnarray}
{J_4}&=&16[(Q_{1R}^2-Q_{1L}^2)(P_{2R}^2-P_{2L}^2)-4P_{2R}^2q^2]\cr
&=&16^2[Q_1Q_2P_1P_2-{\textstyle{1\over
4}}q^2(P_1+P_2)^2] \ .
\label{fourfivepar}
\end{eqnarray}
Comparing with (\ref{4ent}), we learn that for the five-parameter generating
solution the BH  entropy
is given by
\begin{equation}
S_{BH}={\pi\over 8}\sqrt{J_4}.
\label{4Jent}
\end{equation}
This  result generalises the one in   \cite{KK},  where the result for the
four-parameter solution with
$q=0$ was established.

Acting on the generating solution with $SU(8)$ transformations  to generate the
general
charged black hole, and
then with a $E_7$ transformation to generate the solution with  general
asymptotic values of scalar fields, leaves the
BH entropy  (\ref{4Jent}) invariant,
since $J_4$ is invariant. As the dressing of the charges is by an $E_7$
transformation, i.e.  ${\bar {\cal Z}}
={{\cal V}_\infty {\cal Z}}$, the dependence on the asymptotic values of
scalar fields  ${\cal V}_\infty$ drops out of the BH
entropy, which  thus can be
written in terms of the  bare charges alone, as expected.

\subsubsection{D=5}

The BH entropy  of the three-parameter static BPS-saturated generating
solution is
\cite{VS,TMpl,CY5r}:
\begin{equation}
S_{BH}= 2\pi\sqrt{Q_1Q_2\tilde Q} \ .
\label{5ent}
\end{equation}
The  cubic  $E_{6(6)}$
  invariant    $J_3$ constructed from the charge matrix $Z_{5\,AB}$ is
\cite{Cremmer}:
\begin{equation}
{J_3}=-\sum_{A,B,C,D,E,F=1}^{8}\Omega^{AB} Z_{5\,BC} \Omega^{CD} Z_{5\,DE}
\Omega^{EF} Z_{5\,FA}   \ .
\label{cubic}
\end{equation}
By transforming  $Z_5$ to  a skew-diagonal   matrix $Z^0_{5}$, given in terms
of
 the four constrained real eigenvalues $\lambda_i$
(\ref{threepar}),
 and using
(\ref{explo})
for  $\Omega_{AB}$, the
 cubic  form $J_3$ can be written as
\begin{equation}
{J_3}=2\sum_{i=1}^4 \lambda_i^3.
\label{diagfive}
\end{equation}
For the (three  charge  parameter) generating  solution the eigenvalues are
  (\ref{threepar}) so that  (\ref{diagfive}) is:
\begin{eqnarray}
{J_3}&=&12(Q_{1R}^2-Q_{1L}^2){\tilde Q}\cr
&=&48Q_1Q_2\tilde Q,
\label{Jfivegen}
 \end{eqnarray}
which together with (\ref{5ent}) implies
\begin{equation}
S_{BH}=\pi\sqrt{{\textstyle{1\over {12}}}J_3}.
\label{5Jent}
\end{equation}
This  result gives the entropy in terms of the cubic invariant for the
generating
solution, and so, by $U$-invariance,
for all charged static BPS-saturated $D=5$ black holes, as
conjectured in \cite{FK,Pol,CHU}.

\subsubsection{$6 \le D \le 9$}

A BPS-saturated black hole in $D\ge 6$ dimensions should have
a horizon area that is an invariant under $U$-duality constructed from the bare
charges alone,
involving no scalars. This would involve, for example,  constructing an
$SO(5,5)$ singlet from tensor
products of charges transforming as a chiral spinor {\bf 16} of  $SO(5,5)$ in
$D=6$ and
constructing singlets of $SL(5,{\bf R})$ from tensor products of  charges
transforming as a {\bf 10}.
There are no such non-trivial singlets in $D\ge 6$, so that the only invariant
result for the area
is zero, which is precisely what is found (see Appendix B).
Indeed, the  generating solution  for   BPS-saturated solution in $D\ge 6$ has
{\it zero} horizon area \cite{CYNear} and so  zero BH entropy.

\subsubsection{Entropy of Non-Extreme Black Holes}

We now comment on the BH entropy of non-extreme black holes, in particular,
static  black holes in $6\le D\le 9$. (For another approach  to address the BH
entropy of non-extreme black holes, where  additional auxiliary charges are
introduced, see \cite{HMS,KR}.)
The  non-extreme generating solutions are specified in terms of two electric
charges $Q_{1,2}$ and a parameter
$m$ which
measures the deviation from extremality,  i.e.  the BPS-saturated limit is
reached when $m=0$ while the charges $Q_{1,2}$ are kept constant (see Appendix
B).
The BH entropy is given by (\ref{DBHent}) in Appendix B,
which in the  near-BPS-saturated limit ($Q_{1,2}\gg m$) reduces to
the form (\ref{DBPSBHent}):
\begin{equation}
S_{BH}=4\pi\sqrt{{\textstyle{1\over{(D-3)^2}}}Q_1Q_2{(2m)}^{2\over{D-3}}}.
\label{Dent}
\end{equation}

The  BH entropy of the non-extreme black holes  can also be rewritten in a
manifestly
duality-invariant manner. We demonstrate this for  static near-extreme black
holes in $D=7$; examples  of such black holes in other  dimensions    are
similar.
The $5 \times 5$ matrix of dressed charges $Z_{7ij}$ transforms as an
antisymmetric tensor under the local $SO(5)$ symmetry but is invariant under
the rigid
$SL(5,{\bf R})$ duality symmetry. (The central charges are given by the $4
\times 4$ matrix
$Z_{7ij} \gamma^{ij}$ where $ \gamma^{ij}$ are the generators of $SO(5)$ in the
4-dimensional spinor representation.)
The two skew eigenvalues $\lambda_i$ of $Z_{7ij} $
are given by
$\lambda_i=Q_{R,L} $ and these are invariant under
$SO(5) \times SL(5,{\bf R})$. The BH entropy in the near-extreme case is then
\begin{equation}
S_{BH}= {{\pi}\over 2}\sqrt{ (|\lambda_1|^2- |\lambda_2|^2){(2m)}^{1\over{2}}}\
{},
\label{Denta}
\end{equation}
which can be rewritten in a  manifestly $U$-duality invariant form as
\begin{equation}
S_{BH}= {{\pi}\over 2} \left(
[2tr(Y_7^2)-{\textstyle {1\over 2}}(tr Y_7)^2]m \right )^{1\over 4} \ .
\label{Dentb}
\end{equation}
Here $Y_7\equiv Z_7^tZ_7$ and we have used the relationship: $tr(Y_7^m)=2
(-|\lambda_1|^2)^{m}+(-|\lambda_2|^2)^{m}$. Note that now  the entropy does
depend on
the asymptotic values of the scalar
fields.

\subsection{ADM Masses and Supersymmetry Breaking}

We now comment on a $U$-duality invariant form of the  black hole mass
 formula  for BPS-saturated black holes with  different  numbers
 of preserved supersymmetries.  We shall derive such expressions in $D=4$.
Examples  in other dimensions are  similar.

As discussed in the beginning of this Section, in $D=4$ the BPS-saturated
black holes  will preserve ${p\over 8}$  of the $N=8$
supersymmetry if the  BPS-saturated ADM mass  $M$
is equal to $|\lambda _1|=\cdots =|\lambda_p|$ where   $\lambda _i$
($i=1,\cdots, 4$) are
eigenvalues  (\ref{diag4})  of central charge matrix  $Z_4$. Without loss of
generality one  can order the eigenvalues in  such a way that  $|\lambda_i|\ge
|\lambda_j|$ for $j\ge i$ .
Note also that  from (\ref{diag4})
\begin{equation}
tr(Y_4^m)= 2\sum _{i=1}^4(-|\lambda_i|^2)^{m}\ ,
\end{equation}
where $Y_4\equiv Z_4^\dagger Z_4$ and $m=1,\cdots, 5-p$.

\subsubsection{p=4}

These solutions  preserve ${1\over 2}$  of $N=8$ supersymmetry and
\begin{equation}
M=|\lambda_1|=|\lambda_2|=|\lambda _3|=|\lambda_4|.
\end{equation}
Examples of  such solutions are obtained from the
generating solution with only
one non-zero charge, e.g.,
those with only $Q_1\ne 0$.
The mass can be written in a $U$-invariant form as
\begin{equation}
M=-{\textstyle{1\over 8}}trY_4\ .
\end{equation}

\subsubsection{p=3}

These solutions preserve ${3\over 8}$ of $N=8$  supersymmetry  thus
\begin{equation}
M=|\lambda_1|=|\lambda_2|=|\lambda _3|>|\lambda_4|\ .
\end{equation}
An example of  such a generating solution
corresponds to the case with  $(Q_1,
 Q_2, P_1=P_2)\ne 0$, while non-zero $q$ is  determined in
terms of other nonzero charges as  $q={\textstyle{1\over
2}}[(Q_{1R}+P_{2R})^2-Q_{1L}^2]^{1/2}$.
The $U$-duality invariant form of the  BPS-saturated mass can now be written in
terms of  two invariants:
\begin{equation}
tr Y_4=-6|\lambda _1|^2-2|\lambda _4|^2, \qquad
tr
(Y_4^2)=
6|\lambda _1|^4+2|\lambda _4|^4
\end{equation}
as the (larger) root of the quadratic equation:
\begin{equation}
48 M^4 +12 tr Y_4 M^2 + (tr Y_4)^2-2 tr(Y_4^2)=0 \ ,
\end{equation}
\begin{equation}
M^2=-{\textstyle{1\over 8}} tr Y_4 + \sqrt{ {\textstyle{1\over {24}} tr (Y_4^2)
- {\textstyle{1\over {192}}}(tr Y_4)^2}}\ .
\end{equation}

\subsubsection{p=2}

These solutions preserve ${1\over 4}$ of $N=8$  supersymmetry and have the mass
\begin{equation}
M=|\lambda_1|=|\lambda_2|>|\lambda _3|\ge|\lambda_4| \ .
\end{equation}
An  example of  such a generating  solution
is the
case with  only  $(Q_1,Q_2)\ne 0$.
The general mass can be written in terms of the three invariants $tr (Y_4^m)$
for $m=1,2,3$.
The $U$-duality invariant expression for the BPS-saturated mass formula is then
given by   the  (largest) root of a cubic equation; we do not give it
explicitly
here.
Some simplification is obtained if $|\lambda _3|=|\lambda_4|$, in which case
only two invariants
and a quadratic equation are needed.

\subsubsection{p=1}

These solutions preserve ${1\over 8}$ of the $N=8$  supersymmetry and the mass
is
\begin{equation}
M=|\lambda_1|> |\lambda_2|\ge|\lambda _3|\ge|\lambda_4|\ .
\end{equation}
Examples of such    generating  solutions are the
case with  only
$(Q_1,Q_2,P_1)\ne 0$ and  the generating solution with all the five charges
non-zero
is also  in this class.
The  $U$-invariant mass can be written in terms of the four invariants $tr
(Y_4^m)$
for $m=1,2,3,4$ and  involves the (largest) root of a quartic equation
so we do not give it explicitly
here.

\acknowledgments
We would like to thank  K. Chan, S. Gubser, F. Larsen,  A. Sen,  A. Tseytlin
and  D. Youm for useful
discussions.
The work is supported
by the Institute for Advanced Study funds and J. Seward Johnson
foundation (M.C.), U.S. DOE Grant No. DOE-EY-76-02-3071 (M.C.), the NATO
collaborative research grant CGR No. 940870 (M.C.) and the National
Science Foundation Career Advancement Award No. PHY95-12732
(M.C.).
The authors  acknowledge the hospitality of  the Institute for Theoretical
Physics  at the University
of California, Santa Barbara,  where the work was initiated,  the
hospitality of the Department
 of Applied Mathematics and Theoretical Physics  of  Cambridge (M.C.) and of
the CERN Theory Division (M.C.).

\vskip2.mm
\newpage
\section{Appendix A: Effective Action of the NS-NS
Sector of Type II String on Tori}

For the sake of completeness we briefly summarize the form of
the  effective action of the  NS-NS sector for  the toroidally compactified
Type II string in $D$-dimensions (see, e.g., \cite{MS}.). The notation used is
that of \cite{CY4r}.

The compactification of the  $(10-D)$-spatial coordinates
on a $(10-D)$-torus is achieved by choosing the following
abelian Kaluza-Klein Ansatz for the ten-dimensional metric
\begin{equation}
\hat{G}_{MN}=\left(\matrix{e^{a\varphi}g_{{\mu}{\nu}}+
G_{{m}{n}}A^{(1)\,m}_{{\mu}}A^{(1)\,n}_{{\nu}} & A^{(1)\,m}_{{\mu}}
G_{{m}{n}}  \cr  A^{(1)\,n}_{{\nu}}G_{{m}{n}} & G_{{m}{n}}}\right),
\label{4dkk}
\end{equation}
where $A^{(1)\,m}_{\mu}$ ($\mu = 0,1,...,D-1$;
$m=1,...,10-D$) are $D$-dimensional Kaluza-Klein $U(1)$ gauge fields,
$\varphi \equiv \hat{\Phi} - {1\over 2}{\rm ln}\,{\rm det}\, G_{mn}$
is the $D$-dimensional dilaton field, and $a\equiv {2\over{D-2}}$.
Then, the affective action is specified by the following massless
bosonic fields: the (Einstein-frame) graviton $g_{\mu\nu}$, the
dilaton $e^{\varphi}$, $(20-2D)$ $U(1)$ gauge fields ${\cal A}^i_{\mu}
\equiv (A^{(1)\,m}_{\mu},A^{(2)}_{\mu\,m})$ defined
as $A^{(2)}_{\mu\,m} \equiv \hat{B}_{\mu m}+\hat{B}_{mn}
A^{(1)\,n}_{\mu}$,
and the following symmetric $O(10-D,10-D)$
matrix of the scalar fields (moduli):
\begin{equation}
M=\left ( \matrix{G^{-1} & -G^{-1}C  \cr
-C^T G^{-1} & G + C^T G^{-1}C}
\right ),
\label{modulthree}
\end{equation}
where $G \equiv [\hat{G}_{mn}]$, $C \equiv [\hat{B}_{mn}]$ and
 are defined in terms of the
internal parts of ten-dimensional fields.  Then the  NS-NS sector of
the $D$-dimensional effective action takes the form:
\begin{eqnarray}
{\cal L}&=&{1\over{16\pi G_D}}\sqrt{-g}[{\cal R}_g-{1\over (D-2)}
g^{\mu\nu}\partial_{\mu}\varphi\partial_{\nu}\varphi+{1\over 8}
g^{\mu\nu}{\rm Tr}(\partial_{\mu}ML\partial_{\nu}ML)\cr&-&{1\over{12}}
e^{-2a\varphi}g^{\mu\mu^{\prime}}g^{\nu\nu^{\prime}}
g^{\rho\rho^{\prime}}H_{\mu\nu\rho}H_{\mu^{\prime}\nu^{\prime}
\rho^{\prime}}
-{1\over 4}e^{-a\varphi}g^{\mu\mu^{\prime}}g^{\nu\nu^{\prime}}
{\cal F}^{i}_{\mu\nu}(LML)_{ij}
{\cal F}^{j}_{\mu^{\prime}\nu^{\prime}}],
\label{effaction}
\end{eqnarray}
where $g\equiv {\rm det}\,g_{\mu\nu}$, ${\cal R}_g$ is the Ricci
scalar of $g_{\mu\nu}$, and ${\cal F}^i_{\mu\nu} = \partial_{\mu}
{\cal A}^i_{\nu}-\partial_{\nu} {\cal A}^i_{\mu}$ are the
$U(1)^{20-2D}$ gauge field strengths and  $H_{\mu\nu\rho} \equiv
(\partial_{\mu}B_{\nu\rho}-{1\over 2}{\cal A}^i_{\mu}L_{ij}
{\cal F}^j_{\nu\rho}) + {\rm cyc.\ perms.\ of}\  \mu , \nu , \rho$
is the field strength of the two-form field $B_{\mu\nu}$.

The $D$-dimensional effective action (\ref{effaction}) is
invariant under the $O(10-D,10-D)$ transformations ($T$-duality):
\begin{equation}
M \to \Omega M \Omega^T ,\ \ \ {\cal A}^i_{\mu} \to \Omega_{ij}
{\cal A}^j_{\mu}, \ \ \ g_{\mu\nu} \to g_{\mu\nu}, \ \ \
\varphi \to \varphi, \ \ \ B_{\mu\nu} \to B_{\mu\nu},
\label{tdual}
\end{equation}
where $\Omega$ is an $O(10-D,10-D)$ invariant matrix, {\it i.e.},
with the following property:
\begin{equation}
\Omega^T L \Omega = L ,\ \ \ L =\left ( \matrix{0 & I_{10-D}\cr
I_{10-D} & 0 } \right ),
\label{4dL}
\end{equation}
where $I_n$ denotes the $n\times n$ identity matrix.

In $D=4$ the field strength of the abelian gauge
field is self-dual, i.e. $\tilde{\cal F}^{i\,\mu\nu} = {1\over 2\sqrt{-g}}
\varepsilon^{\mu\nu\rho\sigma}{\cal F}^i_{\rho\sigma}$,
 and thus the charged solutions are specified by the
electric and magnetic charges.

In  $D=5$ the effective action is specified by the
graviton, 26 scalar fields (25 moduli fields in the matrix $M$ and
the dilaton $\varphi$), 10 $U(1)$ gauge fields, and the
field strength $H_{\mu\nu\rho}$ of the two-form field $B_{\mu\nu}$.
By the duality transformation $
H^{\mu\nu\rho}=-{e^{4\varphi/3}\over{2!\sqrt{-g}}}
\varepsilon^{\mu\nu\rho\lambda\sigma}{\tilde F}_{\lambda\sigma},$
$H_{\mu\nu\rho}$ can be related to the field strength
$\tilde F_{\mu\nu}$ of the gauge field $ \tilde A_\mu$, which specifies
an additional electric charge $\tilde Q$.

In $D\ge 6$ there the allowed charges are only electric charges
associated with the $(20-2D)$  NS-NS sector abelian gauge fields.

\section{Appendix B: Static Generating Solutions}

For the sake of simplicity we  present explicitly only the
explicit  solution for the non-extreme {\it static} generating solution
in $D=4$,  $D=5$,  $6\le D \le 9$ and  with four\cite{CYI,HLMS}, three
\cite{CY5r,HMS} and two charge\cite{Peet}
parameters of the NS-NS sector, respectively. Note that the full
 generating solution in  $D=4$ is parameterised by
{\it five} charge parameters. For the  explicit form of  the rotating
generating solution  in $D=5$ see \cite{CY5r}  and in  $6\le D\le 9$ see
\cite{CYNear,Llatas}, while  in  $D=4$ the four charge parameter
  rotating solution  is
given in \cite{CY4r} and the five charge static solution is given in
\cite{CY5r}. The parameterisation there is given in terms of  the
``toroidal'' sector of toroidally compactified heterotic string.

We choose to parameterise the generating solutions in terms of the mass
$m$ of the $D$-dimensional Schwarzschild solution, and the boost
parameters $\delta_i$, specifying the charges of the solution.
The notation used is similar to that in \cite{CTIII}.

\subsection{$D=4$-Four Charge Static Solution}
The expression for  the non-extreme dyonic rotating
black hole solution in terms of the (non-trivial)
four-dimensional bosonic fields is of the following form
\footnote{The four-dimensional Newton's constant is taken to be
$G_N^{D=4}={1\over 8}$ and we follow the convention of \cite{MP}, for the
definitions of the ADM mass, charges,
dipole moments and angular momenta.}:
\begin{eqnarray}
ds^2_{E}&=& -\lambda fdt^2+\lambda^{-1}[f^{-1}dr^2+r^2d\Omega_2^2
],\cr
G_{11}&=&{{T_1}\over{T_2}}, \ \
G_{22}={{F_2}\over {F_1}}, \ \
e^{2\varphi}={{F_1F_2}\over{T_1T_2}},
\label{4dsol}
\end{eqnarray}
where $ds^2_{E}$ specifies the  Einstein-frame ($D$-dimensional) space-time
line element,  $G_{ij}$ correspond to the internal toroidal metric coefficients
 and $\varphi$ is the $D$-dimensional dilaton field (see Appendix A). Other
scalar fields are constant and assume canonical values (one or zero).
Here
\begin{equation}
f=1-{{2m}\over r}, \ \ \lambda=(T_1T_2F_1F_2)^{-{1\over 2}}
\label{4fl}
\end{equation}
and
\begin{eqnarray}
T_1&=&1+{{2m{\rm sinh}^2 \delta_{e1}}\over  r},\ \
T_2=1+{{2m{\rm sinh}^2 \delta_{e2}}\over  r}, \cr
F_1&=&1+{{2m{\rm sinh}^2 \delta_{m1}}\over  r}, \ \
F_2=1+{{2m{\rm sinh}^2 \delta_{m2}}\over  r},
\label{4dpar}
\end{eqnarray}
The ADM mass,  and four $U(1)$ charges $Q_1,Q_2,P_1,
P_2$, associated with the respective gauge fields
$A^{(1)}_{1\mu},A^{(2)}_{1\mu},A^{(1)}_{2\mu},A^{(2)}_{2\mu}$,  can be
expressed
in terms of $m$, and four boosts $\delta_{e1,e2,m1,m2}$ in the following way:
\begin{eqnarray}
M&=&4m({\rm cosh}^2 \delta_{e1}+{\rm cosh}^2 \delta_{e2}+
{\rm cosh}^2 \delta_{m1}+{\rm cosh}^2 \delta_{m2})-8m, \cr
Q_1&=&4m{\rm cosh}\delta_{e1}{\rm sinh}\delta_{e1},\ \ \ \ \
Q_2= 4m{\rm cosh}\delta_{e2}{\rm sinh}\delta_{e2}, \cr
P_1&=&4m{\rm cosh}\delta_{m1}{\rm sinh}\delta_{m1},\ \ \ \ \
 P_2 =4m{\rm cosh}\delta_{m2}{\rm sinh}\delta_{m2}.
\label{4dphys}
\end{eqnarray}
The BH entropy is of the form\cite{CYI,HLMS}:
\begin{equation}
S_{BH}=2\pi
(4m)^2\cosh\delta_{e1}\cosh\delta_{e2}\cosh\delta_{m1}\cosh\delta_{m2},
\label{4BHent}
\end{equation}
which in the  BPS-saturated limit ($m\to 0$, $\delta_{e1,e2,m1,m2}\to
\infty$ while keeping $Q_{1,2}$, $P_{1,2}$ finite) reduces to
the form \cite{CY}:
\begin{equation}
S_{BH}=2\pi\sqrt{Q_1Q_2P_1P_2}.
\label{4BPSBHent}
\end{equation}
In the case of the fifth charge parameter  $q$ added, the BH entropy
of the BPS-saturated   black holes becomes \cite{CTII}:
\begin{equation}
S_{BH}=2\pi\sqrt{Q_1Q_2P_1P_2-{\textstyle{1\over 4}}q^2(P_1+P_2)^2}.
\label{4BPSBH5p}
\end{equation}

\subsection{$D=5$-Three Charge Static Solution}
The expression for  the non-extreme dyonic rotating
black hole solution in terms of the (non-trivial)
five-dimensional bosonic fields is of the following form
\cite{CY5r,HMS}\footnote{The five-dimensional Newton's constant is taken to be
$G_N^{D=5}={{2\pi}\over 8}$.}:
\begin{eqnarray}
ds^2_{E}&=& -\lambda^2 fdt^2+\lambda^{-1}[f^{-1}dr^2+r^2d\Omega_3^2
],\cr
G_{11}&=&{{T_1}\over{T_2}}, \ \
e^{2\varphi}={\tilde T^2\over{T_1T_2}},
\label{5dsol}
\end{eqnarray}
with other scalars  assuming  constant canonical values.
Here
\begin{equation}
f=1-{{2m}\over r^2}, \ \ \lambda=(T_1T_2\tilde T)^{-{1\over 3}},
\label{5fl}
\end{equation}
and
\begin{eqnarray}
T_1&=&1+{{2m{\rm sinh}^2 \delta_{e1}}\over  r^2}, \ \
T_2=1+{{2m{\rm sinh}^2 \delta_{e2}}\over  r^2}, \ \
\tilde T=1+{{2m{\rm sinh}^2 \delta_{\tilde e}}\over  r^2} \ .
\label{5dpar}
\end{eqnarray}
The ADM mass, three charges $Q_{1,2},\tilde Q$  associated with
respective gauge fields $A_{1\mu}^{(1)}, \ A_{1\mu}^{(2)}$
 and ${\tilde A}_\mu$ (the  gauge field related to the  two
from field $B_{\mu\nu}$ by a duality transformation), are expressed
in terms of $m$, and three  boosts $\delta_{e1,e2,\tilde e}$ in the following
way:
\begin{eqnarray}
M&=&2m({\rm cosh}^2 \delta_{e1}+{\rm cosh}^2 \delta_{e2}+
{\rm cosh}^2 \delta_{\tilde e})-3m, \cr
Q_1&=&2m{\rm cosh}\delta_{e1}{\rm sinh}\delta_{e1},\ \
 Q_2= 2m{\rm cosh}\delta_{e2}{\rm sinh}\delta_{e2},\ \
{\tilde Q}=2m{\rm cosh}\delta_{\tilde e}{\rm sinh}\delta_{\tilde e}.
\label{5dphys}
\end{eqnarray}
The BH entropy is of the form \cite{HMS}:
\begin{equation}
S_{BH}=2\pi (2m)^{3\over 2}\cosh\delta_{e1}\cosh\delta_{e2}\cosh\delta_{\tilde
e}
\label{5BHent},
\end{equation}
which in the BPS-saturated limit ($m\to 0$, $\delta_{e1,e2,\tilde e}\to
\infty$ with $Q_{1,2}, \ \tilde Q$ finite) reduces to
the form \cite{VS,TMpl}:
\begin{equation}
S_{BH}=2\pi\sqrt{Q_1Q_2\tilde Q}.
\label{5BPSBHent}
\end{equation}

\subsection{$6\le D\le 9$-Two Charge Static Solution}

The expression for  the non-extreme dyonic rotating
black hole solution in terms of the (non-trivial)
five-dimensional bosonic fields is of the following form
\cite{Peet,HSen,CYNear,Llatas}\footnote{The $D$-dimensional Newton's constant
is taken to be
$G_N^{D}={{(2\pi)^{D-4}}\over 8}$.}:
\begin{eqnarray}
ds^2_{E}&=& -\lambda^{D-3} fdt^2+\lambda^{-1}[f^{-1}dr^2+r^2d\Omega_{D-2}^2
],\cr
G_{11}&=&{{T_1}\over{T_2}}, \ \
e^{2\varphi}={1\over{T_1T_2}},
\label{Ddsol}
\end{eqnarray}
while other scalar fields are constant and  assume  canonical values.
Here
\begin{equation}
f=1-{{2m}\over r^{D-3}}, \ \ \lambda=(T_1T_2)^{1\over{D-2}},
\label{Dfl}
\end{equation}
and
\begin{equation}
T_1=1+{{2m{\rm sinh}^2 \delta_{e1}}\over  r^{D-3}}, \ \
T_2=1+{{2m{\rm sinh}^2 \delta_{e2}}\over  r^{D-3}}.
\label{Ddpar}
\end{equation}
The ADM mass, $U(1)$ charges $Q_1,Q_2$  associated with
respective
$A_{1\mu}^{(1)}, \ A_{2\mu}^{(2)}$ gauge fields, are expressed
in terms of $m$, and two  boosts $\delta_{e1,e2}$
 in the following way:
\begin{eqnarray}
M&=&{{\omega_{D-2}m}\over{8\pi G_D}}[ (D-3)({\rm cosh}^2( \delta_{e1}+{\rm
cosh}^2 \delta_{e2})-(D-4)], \cr
Q_1&=&{{\omega_{D-2}m}\over{8\pi G_D}}(D-3){\rm cosh}\delta_{e1}{\rm
sinh}\delta_{e1},\
 Q_2={{\omega_{D-2}m}\over{8\pi G_D}}(D-3){\rm cosh}\delta_{e2}{\rm
sinh}\delta_{e2},\ \
\label{Ddphys}
\end{eqnarray}
where $\omega_{D-2}={{2\pi^{{D-1}\over 2}}/ {\Gamma({{D-1}\over 2})}}$.

The BH entropy is of the form:
\begin{equation}
S_{BH}={{\omega_{D-2}}\over {2G_D}}
m^{{D-2}\over{D-3}}\cosh\delta_{e1}\cosh\delta_{e2},
\label{DBHent}
\end{equation}
which in the  near-BPS-saturated limit ($Q_{1,2}\gg m$) reduces to
the form:
\begin{equation}
S_{BH}=4\pi\sqrt{{\textstyle{1\over{(D-3)^2}}}Q_1Q_2{(2m)}^{2\over{D-3}}}.
\label{DBPSBHent}
\end{equation}

\begin{table}
\begin{tabular}{|l|c|c|c|}
D& Classical Duality
-${ U}$ &
Maximal Compact Subgroup-${ C_U}$& Quantum Duality-${Q_U}$ \\
\hline
4 & $E_{{7(7)}}$&$SU(8)$ &$E_{{7(7)}}({\bf Z})$ \\
5 & $E_{{6(6)}}$&$USp(8)$ &$E_{{6(6)}}({\bf Z})$ \\
6 & $SO(5,5)$&$SO(5)\times SO(5)$ &$SO(5,5;{\bf Z})$ \\
7 & $SL(5,{\bf R})$&$SO(5)$ &$SL(5, {\bf Z})$ \\
8 & $SL(3,{\bf R})\times SL(2, {\bf R})$&$SO(3)\times U(1)$ &$SL(3, {\bf
Z})\times
SL(2,{\bf Z})$ \\
9 & $SL(2,{\bf R})\times {\bf R}^+$&$ U(1)$ &$SL(2, {\bf Z}) $\\
\end{tabular}
\caption[]{The classical   and   quantum   duality
symmetries \cite{HT}  for toroidally
 compactified  Type  II string in $ 4\le D\le 9$. }
\end{table}

\end{document}